\documentclass[preprint]{aastex7}

\begin{document}

\title{Asteroseismic mass and radius of the naked-eye red giant HD145250}

\author[0000-0002-8159-1599]{L\'aszl\'o Moln\'ar}
\affiliation{HUN-REN CSFK, Konkoly Observatory, Konkoly Thege Mikl\'os \'ut 15-17, Budapest, 1121 Hungary}
\affiliation{E\"otv\"os Lor\'and University, Institute of Physics and Astronomy, P\'azm\'any P\'eter s\'et\'any 1, Budapest, Hungary}
\email{molnar.laszlo@csfk.org}

\author[0009-0007-3760-515X]{Kl\'ara Lelkes}
\affiliation{HUN-REN CSFK, Konkoly Observatory, Konkoly Thege Mikl\'os \'ut 15-17, Budapest, 1121 Hungary}
\affiliation{E\"otv\"os Lor\'and University, Institute of Physics and Astronomy, P\'azm\'any P\'eter s\'et\'any 1, Budapest, Hungary}
\email{lelkes.klara@csfk.org}

\correspondingauthor{L\'aszl\'o Moln\'ar (molnar.laszlo@csfk.org)}


\begin{abstract}

We present the first asteroseismic analysis of the bright, nearby red giant star, HD145250. We calculate the global seismic quantities of the star from single-sector, 2-minute TESS photometry, and determine its mass and radius to be $\sim1.4$\,M$_\odot$ and $\sim16$\,R$_\odot$ using asteroseismic scaling relations. Our values agree with published non-seismic mass and radius estimates based on comparisons with stellar evolutionary models.

\end{abstract}

\keywords{\uat{Stellar astronomy}{1583} --- \uat{Asteroseismology}{73} --- \uat{Stellar photometry}{1620}}


\section{Introduction}

HD145250 is a naked-eye $(V=5.1\,{\rm mag})$ red giant in the constellation Scorpion, found in between the Sun and the Upper Sco association behind it. Despite its brightness and position, it was not observed during the K2 mission of the Kepler space telescope, and it has not included in any TESS surveys \citep[e.g., in][]{Hon-2021}, and it is featured in a relatively low number of studies. 

HD145250 was considered a single star by \citet{Eggleton_2008MNRAS.389..869E} who found no significant evidence for multiplicity based on radial-velocity measurements. However, \citet{Kervella_2019A&A...623A..72K} suggest the possible presence of a faint companion based on proper motion anomaly derived from comparison of \textit{Hipparcos} and \textit{Gaia} DR2 data. Here we present the first asteroseismic analysis of the star, based on asteroseismic scaling relations \citep{Kjeldsen-Bedding-1995,Huber-2011}. 

\section{Data and methods}

\begin{figure*}
\centering
\includegraphics[width=0.99\textwidth]{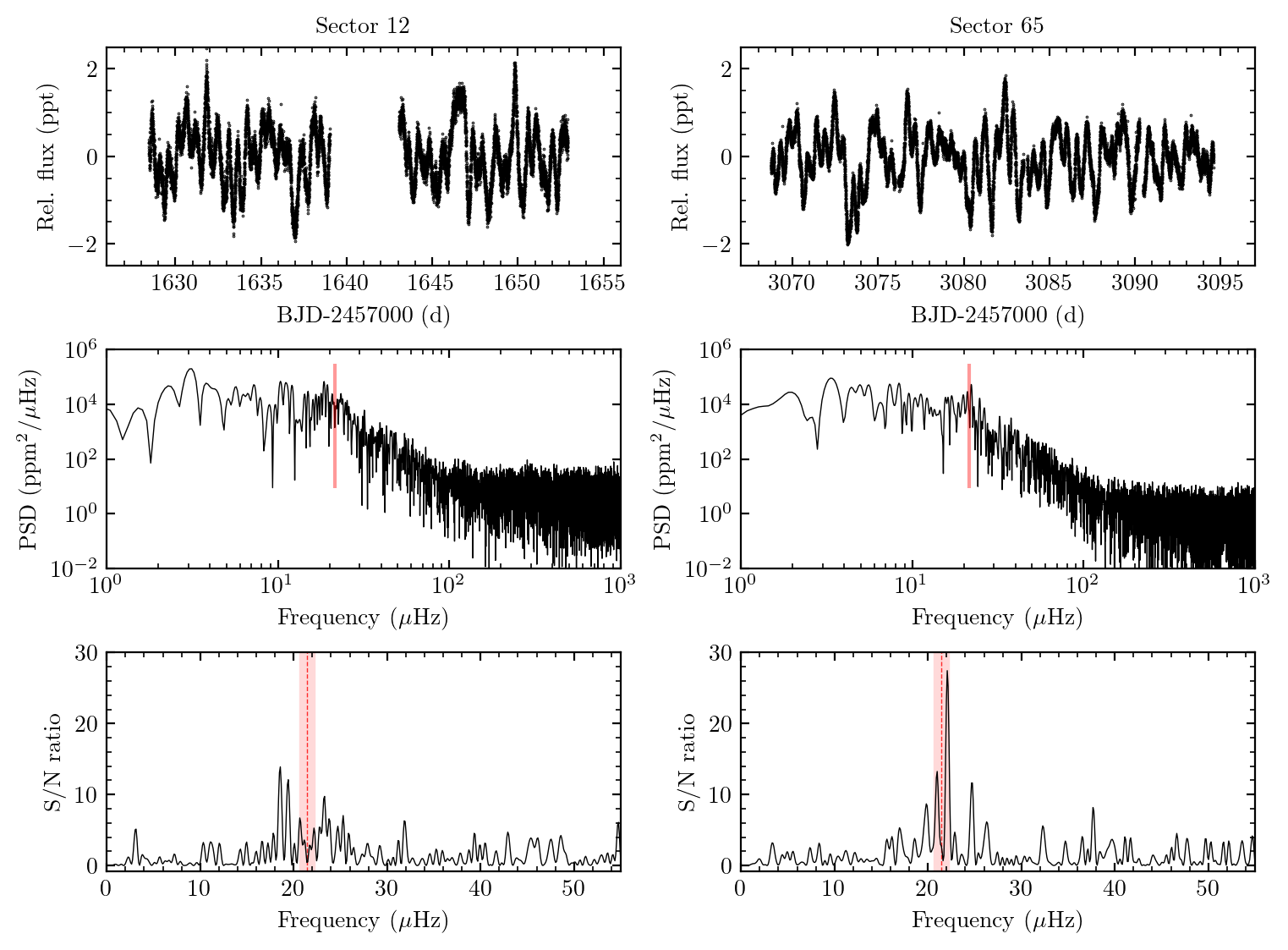}
\caption{Top: TESS 2-minute cadence photometry of HD145250. Middle: power density spectra of the light curves. Bottom: oscillation signals in the data after removal of the granulation background. Red bands indicate $\nu_{\rm max}$ values determined for S65.
\label{figure}}
\end{figure*}

The TESS space telescope \citep{Ricker-2015} observed the star in 2019 and 2023, during Sectors 12 and 65, and  on both occasions in 2-minute short cadence mode. We carried out our initial analysis of the PDCSAP (Pre-search data conditioned simple aperture photometry) data with the \texttt{lightkurve} software tool \citep{lightkurve-2018}. We found that the star displays clear photometric variability caused by acoustic oscillations, and we identified the corresponding power excess in the power density spectrum. However, these appear at quite low frequencies, making the individual modes unresolved at the short time spans of single TESS sectors, as we show in Fig.~\ref{figure}. 

We then used the \texttt{pySYD} software to calculate the global asteroseismic quantities of the star \citep{Chontos-2022}. Data from Sector 65 provided a clear detection of the power excess with multiple frequency peaks, and we determined the frequency of maximum oscillation amplitude ($\nu_{\rm max}$) and the large frequency separation ($\Delta \nu$) to be $\nu_{\rm max} = 21.4\pm 1.0\, \mu{\rm Hz}$ and $\Delta \nu = 2.53 \pm 0.20\, \mu{\rm Hz}$, respectively. As Fig.~\ref{figure} shows, the power excess is much less pronounced in S12. That sector gave a much poorer fit which we decided not to use in our analysis. 


Scaling relations require further stellar physical parameters to calculate the mass of the star. We adopted the stellar atmospheric parameters from \cite{Soubiran_2022A&A...663A...4S} who compared $[\text{Fe/H}]$ determinations from the largest spectroscopic surveys with values from reference catalogs. For HD145250, they report $T_{\text{eff}} = 4540 \pm 50\, \text{K}$, $\log g = 2.74 \pm 0.1 $ and $[\text{Fe/H}] = -0.36 \pm 0.05$.
A distance of $86.686 \pm 0.55$\,pc was derived by \citet{Bailer-Jones-2021} using Gaia EDR3 parallax with geometric Galactic priors. 


We calculated the luminosity using the \textit{Gaia} DR3 \textit{G}--band 
brightness ($G = 4.7489 \pm 0.0028$ mag) and a bolometric correction of $BC = -0.158 \pm 0.020$\,mag, following the method provided by \citet{Creevey-2023}\footnote{\url{https://gitlab.oca.eu/ordenovic/gaiadr3\_bcg}}. 
Given the closeness of the star we assumed no interstellar extinction, but incorporated an uncertainty of $\sigma_{A_G} = 0.01$\,mag. From these we derived a bolometric magnitude of $M_{\text{bol}} = -0.099 \pm 0.026$\,mag 
and luminosity of $ L_{\text{bol}} = 86.191 \pm 2.097$\,L$_\odot$. These physical parameters place the star close to, but outside of the He-burning red clump, therefore we assume that it is a H-shell burning red giant star.

Given the uncertainty in $\Delta \nu$, we calculated the mass using two asteroseismic scaling relations:
\begin{equation}
    \frac{M}{M_{\odot}} \simeq \left ( \frac{\nu_{\text{max}}}{f_{\nu_{\text{max}}}\nu_{\text{max},\odot}} \right ) ^{3} \left ( \frac{  \Delta \nu}{f_{\Delta \nu} \Delta \nu_{\odot}} \right ) ^{-4} \left ( \frac{T_{\text{eff}}}{T_{\text{eff},\odot}} \right ) ^{3/2}
    \label{equ:mass_deltanu}
\end{equation}


\begin{equation}
    \frac{M}{M_{\odot}} \simeq \left ( \frac{\nu_{\text{max}}}{f_{\nu_{\text{max}}}\nu_{\text{max},\odot}} \right ) \left ( \frac{L}{L_{\odot}} \right ) \left ( \frac{T_{\text{eff}}}{T_{\text{eff},\odot}} \right ) ^{-7/2}
    \label{equ:mass_luminosity}
\end{equation}


We can also calculate the radius as the star as:

\begin{equation}
    \frac{R}{R_{\odot}} \simeq \left ( \frac{\nu_{\text{max}}}{f_{\nu_{\text{max}}}\nu_{\text{max},\odot}} \right )  \left ( \frac{ \Delta \nu}{f_{\Delta \nu} \Delta \nu_{\odot}} \right ) ^{-2} \left ( \frac{T_{\text{eff}}}{T_{\text{eff},\odot}} \right ) ^{1/2}
    \label{equ:radius}
\end{equation}

where $f_{\nu_{\text{max}}}$ and $f_{\Delta \nu}$ are the correction factors to the solar scaling. The correction factor $f_{\nu_{\text{max}}}$ was set to 1 \citep{Reyes_2025MNRAS.538.1720R}, and we assumed $f_{\Delta \nu}$ to be 0.97 based on Figure 4 in \cite{Sharma-2016}. Solar values were set to $\nu_{\rm max,\odot} = 3090\pm30\,\mu{\rm Hz}$ and $\Delta \nu_{\odot} = 135.1\pm0.1\,\mu{\rm Hz}$, respectively \citep{Huber-2011}.


\section{Results}

According to Eq.~(\ref{equ:mass_deltanu}), the estimated mass of
HD145250 is $M = 1.67 \pm  0.58$\,M$_\odot$, while Eq.~(\ref{equ:mass_luminosity}) gives a lower estimate of $M = 1.38 \pm 0.09$\,M$_\odot$. The two estimates agree within
the higher uncertainty of Eq.~(\ref{equ:mass_deltanu}), the $\Delta \nu$-based relation. The difference may be due to a possible faint companion or uncertainties in the determination of the $\Delta \nu$ value, which could result from the limited time series. The stellar mass was estimated by \citet{Charbonnel-2020} based on a comparison of its position on the HRD to stellar evolutionary tracks. They found a mass of $1.5^{+1.0}_{-0.4}$\,M$_\odot$, which agrees with our result. 

The seismic stellar radius derived from equation~(\ref{equ:radius}) is 
$R =  16.5 \pm 2.7$ $R_\odot$. For comparison, 
estimates based on stellar models and \textit{Gaia} DR2 photometry
provided a radius of $R = 15.97^{+0.29}_{-1.15}$\,R$_\odot$ \citep{Andrae-2018}. Updated results based on \textit{Gaia} DR3 photometry and high-resolution RVS spectroscopy suggest $R = 17.48\pm0.37$\,R$_\odot$, but at a luminosity of $L = 118.6\pm1.8$\,L$_\odot$, which is significantly higher than our result \citep{Fouseneau-2023}. 

Overall, we confirm via asteroseismology that HD145250, a naked-eye star within 100~pc, and thus among the Gaia Nearby Star sample \citep{GaiaNSC-2021}, is more massive than the Sun and it is currently ascending on the red giant branch.



\begin{acknowledgments}
This research was supported by the `SeismoLab' KKP-137523 grant and the TKP2021-NKTA-64 excellence grant of the Hungarian  Research, Development and Innovation Office (NKFIH). K.L.~acknowledges the undergraduate research assistant program of Konkoly Observatory.
\end{acknowledgments}

\facilities{TESS \citep{Ricker-2015}}
\software{lightkurve \citep{lightkurve-2018},  
          pySYD \citep{pysyd-2021,Chontos-2022}
          }






\bibliography{HD145250}{}
\bibliographystyle{aasjournalv7}



\end{document}